\documentclass[aps,pra,reprint,superscriptaddress,floatfix,showkeys]{revtex4-2}

\usepackage{graphicx}
\usepackage{upgreek}
\usepackage{fixmath}
\usepackage{amsmath}
\usepackage{textcomp}
\usepackage{color}

\begin{document}

\title{On-chip squeezed light in the audio frequency band}

\author{Xuezhi Zhu}
\author{Yunyun Cao}
\author{Rui Liu}
\affiliation{State Key Laboratory of Quantum Optics Technologies and Devices, Institute of Opto-Electronics, Shanxi University, Taiyuan 030006, China}

\author{Yaya He}
\affiliation{National Laboratory of Solid State Microstructures, College of Engineering and Applied Sciences, School of Physics, Collaborative Innovation Center of Advanced Microstructures, Nanjing University, Nanjing 210093, China}

\author{Fan Zhang}
\affiliation{National Laboratory of Solid State Microstructures, College of Engineering and Applied Sciences, School of Physics, Collaborative Innovation Center of Advanced Microstructures, Nanjing University, Nanjing 210093, China}

\author{Yaqing Zhang}
\author{Shiwei Du}
\affiliation{State Key Laboratory of Quantum Optics Technologies and Devices, Institute of Opto-Electronics, Shanxi University, Taiyuan 030006, China}

\author{Meihong Wang}
\affiliation{State Key Laboratory of Quantum Optics Technologies and Devices, Institute of Opto-Electronics, Shanxi University, Taiyuan 030006, China}
\affiliation{Collaborative Innovation Center of Extreme Optics, Shanxi University, Taiyuan 030006, China}

\author{Xiaoshun Jiang}
\affiliation{National Laboratory of Solid State Microstructures, College of Engineering and Applied Sciences, School of Physics, Collaborative Innovation Center of Advanced Microstructures, Nanjing University, Nanjing 210093, China}

\author{Xiaolong Su}
\email{suxl@sxu.edu.cn}
\affiliation{State Key Laboratory of Quantum Optics Technologies and Devices, Institute of Opto-Electronics, Shanxi University, Taiyuan 030006, China}
\affiliation{Collaborative Innovation Center of Extreme Optics, Shanxi University, Taiyuan 030006, China}

\begin{abstract}
Squeezed light in the audio-frequency band is a key resource for quantum metrology and quantum sensing. However, realizing stable audio-frequency squeezed light on integrated photonic platforms remains challenging due to technical noise and the difficulty of scalable phase referencing. Here, we demonstrate on-chip generation of audio-band two-mode squeezed states down to 60 Hz in a silica microcavity. To enable phase-stable operation without directly locking fragile quantum modes, we develop a coherent-comb control method in which a weak electro-optic reference comb co-propagates with the vacuum at the quantum frequency modes in an orthogonal polarization. This scheme provides quadrature measurement and long-timescale phase stability, thereby enabling covariance-matrix reconstruction. We verify the entanglement with the positive partial transposition criterion, which confirms inseparability via a minimum symplectic eigenvalue of 0.395 ($<0.5$). Our results establish an experimentally accessible route toward on-chip phase-stable audio-band squeezing and support the scalable framework for continuous-variable quantum information processing with integrated photonics.

\end{abstract}


\maketitle

\section{Introduction}\label{sec:intro}

Most of relevant signals in precision measurement and sensing reside in the audio-frequency band, notably in gravitational-wave detection and radiation-pressure-limited readout \cite{2019_squ_gravitational-wave_1,2019_squ_gravitational-wave_2,2021_gravitational_wave_review,2024_LIGO_squ_exp,2020_LIGO_radiation_pressure}. 
In many precision sensing scenarios, measurement sensitivity is limited by the shot-noise level (SNL), motivating the use of squeezed light to surpass classical performance \cite{1981_Earliest_QMeasure_theo,2019_sensing_squ_review}. 
Beyond precision sensing \cite{2021_Raman_microscopes,2024_Adaptive+optics_microscopes}, quantum noise reduction plays a central role across a broad range of photonic quantum technologies, including quantum computation \cite{2025_architecture_Xanadu,2025_PsiQ_architecture}, secure communication \cite{2025_jinan1st_qkd,2026_fiber_qdc}, and quantum-enhanced metrology \cite{2026_metrology_atom_exp,2023_metrology_SMSS_bulk,2020_metrology_photonic_review}.
These applications rely on precise control of optical quadratures and phase-sensitive measurements \cite{2024_integratedCV_communication+metrology_review}, motivating the development of integrated platforms capable of delivering stable nonclassical light at low frequencies. 
Although bulk experiments have demonstrated audio-band squeezing \cite{2024_mHz_squ_bulk,2023_10Hz_Squ_cesium_bulk,2025_kHz_brightSqu_bulk}, integrated photonic implementations have largely remained confined to higher sideband frequencies, leaving chip-scale quantum noise reduction at audio frequencies an outstanding challenge. 
In particular, operation in the audio band requires maintaining a SNL measurement baseline in the presence of slow technical noise, detuning fluctuations, and long-timescale phase drift, which can rotate the measurement quadrature and obscure nonclassical correlations even when intracavity squeezing is present.

Optical microcavities provide an attractive route toward chip-scale quantum photonics by combining strong resonant enhancement, high optical quality factors, and compatibility with foundry fabrication, offering a platform in which various quantum resources can be implemented in scalable architectures \cite{2026_silica+CMOS_vahala,2025_vahala_DV_SiN_PhotonPair}. 
At the same time, microcavity-based platforms have enabled highly sensitive measurements across a broad range of physical observables \cite{2015_sensing_WGM_review,2020_sensing_WGM_review,2021_sensing_WGM_review}, including magnetic fields \cite{2020_magnetometers_microcavity_exp}, ultrasound \cite{2020_Optomechanical_microcavity_exp,2024_Ultrasound_sening_review}, and nanoscale particles \cite{2011_sensing_nanoparticle,2014_sensing_nanoparticle}, and provide a natural platform for further quantum-enhanced sensing \cite{2018_magnetometry_sensing_silicon,2024_optomechanics_microfilm_cavity_exp,2025_optomechanics_microFP_exp}. 
Motivated by the maturity of microcavity platforms, microcavity-based continuous-variable (CV) quantum light sources are rapidly advancing as an integrated route for generating squeezed and entangled states \cite{2025_QSource_onChip_review}. Recent advances have demonstrated on-chip generation of nonclassical states including single-mode squeezing \cite{II_NTTWG,II_NTT8dB,II_DuttTunableSqueezing,II_Xanadu_SMSS}, two-mode squeezing \cite{III_Dutt_IntensityDifferenceSS,III_Dutt_HighIntensityDifferenceSS,III_Silica_TMSS,III_Xanadu_TMSS}, and cluster states \cite{2025_cppd_cluster,IV_ClusterGenerationOnSilica} across multiple material platforms to prove applications \cite{I_PSAOnChip,2026_QRNG_CV_chip_exp,I_PhaseSensorOnChip,2022_GBS_Borealis_chip}. However, integrated demonstrations have largely focused on radio-frequency sidebands, leaving a gap for audio-band squeezing between bulk optics and chip-scale implementations, where low-frequency technical drifts and excess noise can dominate.

Achieving stable audio-band squeezing on integrated photonic platforms remains challenging due to technical noise and the difficulty of scalable phase referencing. 
Technical noise that can influence the generation and observation of audio-band squeezing includes parasitic noise associated with nonlinear processes, fluctuations, detuning noise \cite{2025_SMSS_parasitic_nonlinear_processes,2004_audio+squ_detuning+fluctuations}, and residual local-oscillator intensity-noise leakage from imperfect balance \cite{2015_audio+squ_LO+noise,2017_squ_measure_review}. 
In parallel, the difficulty of scalable phase referencing reflects the practical demand for long-term stability of the measurement basis, since slow environmental drift, control-loop feedback noise, and long-term actuator limitations can rotate or wander the effective measurement quadrature. As a result, nonclassical intracavity correlations may not translate into observable audio-band squeezing, rendering them effectively invisible in homodyne measurements over extended timescales \cite{2005_QI_CV_review,2012_Gaussian_QI_review,2012_phase_Yonezawa_Science}. 
Together, these constraints make audio-band squeezing on chip difficult because technical noise can both corrupt the generated state and elevate the low-frequency detection baseline above the SNL, while the slow drift of the quantum state or measurement basis can average away phase-sensitive features over time.

Here, we develop a coherent-comb control (CCC) method for integrated CV photonics that addresses these requirements and enables observable audio-band two-mode squeezing with analysis frequencies down to 60~Hz while maintaining arbitrary quadrature access. In this method, a weak coherent reference chain derived from the pump transfers phase information across multiple frequency modes without applying direct modulation or feedback to the quantum modes themselves, thereby supporting phase-stable quadrature measurements over extended timescales. Using this phase-stable experimental platform, we generate two-mode squeezed states down to 60~Hz in a silica microcavity and verify entanglement via covariance-matrix reconstruction, yielding a minimum symplectic eigenvalue of 0.395 ($<0.5$). Our results provide a pathway toward chip-scale quantum sensors operating in the audio-frequency regime and support integrated CV photonics as a scalable architecture for phase-sensitive quantum information processing.

\section{The principle}\label{sec:model}

In third-order nonlinear materials, quantum frequency modes are commonly generated via four-wave mixing (FWM) in a Kerr-nonlinear microcavity. The Kerr interaction can be written in a unified form as a quartic Hamiltonian \cite{2022_soliton_quantum_theo},
\begin{equation}
\hat H_{\mathrm{Kerr}}
=
\mathrm{i}\hbar
\sum_{u,v,j,k}
g_{uvjk}\,
\delta_{u+v-j-k}
\left(
\hat a_u^\dagger \hat a_v^\dagger \hat a_j \hat a_k
-
\hat a_u \hat a_v \hat a_j^\dagger \hat a_k^\dagger
\right),
\label{eq:Hkerr_unified}
\end{equation}
where $\hat a_m$ is the annihilation operator of cavity mode $m$, $g_{uvjk}$ is the effective Kerr coupling coefficient set by the modal overlap and material nonlinearity, and $\delta_{u+v-j-k}$ represents the mode-matching condition for FWM. Depending on the pump configuration and operating point, Eq.~\eqref{eq:Hkerr_unified} can generate spontaneous pair generation (SPG) of signal--idler sidebands, induce resonance shifts through self- and cross-phase modulation (SPM/XPM) excited by the strong pump field, and cause mode-conversion processes such as FWM-Bragg scattering.

\begin{figure*}[t]
    \centering
    \includegraphics[width=0.85\textwidth]{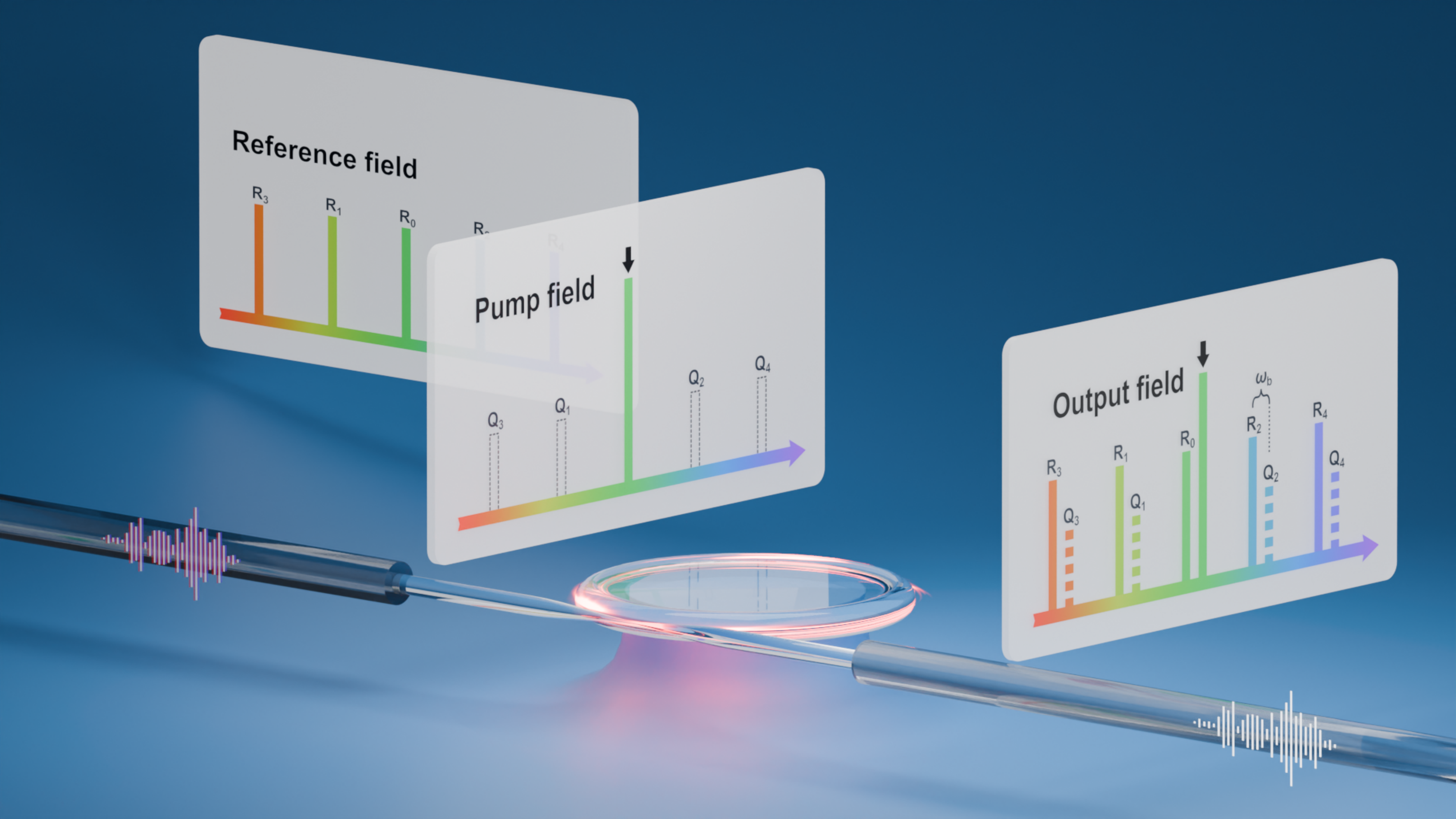}
    \caption{Operating principle of coherent-comb control (CCC) for phase-referenced homodyne detection of microcavity two-mode squeezed states. Paired quantum frequency modes $\mathrm{Q}_i$ are generated via Kerr four-wave mixing, while a weak pump-derived electro-optic reference comb $\mathrm{R}_i$ co-propagates through the same optical path in an orthogonal polarization and with a fixed frequency bias $\omega_{\mathrm{b}}$. After the cavity, the reference and quantum modes are separated and detected independently, so that phase sensing and feedback act on the reference path without introducing a bright locking tone at the quantum signal frequencies.}
    \label{fig:concept}
\end{figure*}

For polychromatic pumping or above threshold, Eq.~\eqref{eq:Hkerr_unified} leads to complex multimode entanglement structures. By contrast, when the microcavity is driven by a monochromatic pump and operated below threshold, SPG dominates and the quantum frequency modes form multiple independent pairs of two-mode squeezed states. In the mean-field approximation of the monochromatic pump, the Hamiltonian becomes quadratic,
\begin{equation}
\begin{split}
\hat H
&=
\mathrm{i}\hbar \sum_{l>0}
\left(
g_{2l}\,\hat a_{2l}^{\dagger}\hat a_{2l+1}^{\dagger}
-
g_{2l}^{*}\,\hat a_{2l}\hat a_{2l+1}
\right)
\\
&\quad+
\hbar \sum_{l>0} \Delta_{2l}^{\mathrm{eff}}
\left(\hat a_{2l}^{\dagger}\hat a_{2l}+\hat a_{2l+1}^{\dagger}\hat a_{2l+1}\right),
\end{split}
\label{eq:Heff_SPWFM}
\end{equation}
where $\hat a_{2l,2l+1}$ are the annihilation operators of the two quantum sideband modes at $\mathrm{Q}_{2l,2l+1}$ (or $\mathrm{Q}_{j}$) in Fig.~\ref{fig:concept}, $g_{2l}$ is the complex SPG-induced parametric coupling rate for the $(\mathrm{Q}_{2l},\mathrm{Q}_{2l+1})$ mode pair, and $\Delta_{2l}^{\mathrm{eff}}$ is the effective detuning of the sidebands, including the static cavity detuning and the SPM/XPM induced resonance shift due to pump light. The first term in Eq.~\eqref{eq:Heff_SPWFM} is the SPG interaction, generating the two-mode squeezed vacuum states. The second term in Eq.~\eqref{eq:Heff_SPWFM} does not generate photons. Instead, it produces a coherent phase rotation of the sideband mode operators at a rate $\Delta_{2l}^{\mathrm{eff}}$, equivalently rotating the squeezing ellipse in phase space. Since the fluctuations of the rotation can wash out phase-sensitive squeezing signatures, the measurement basis needs to be actively locked in practice.

In order to lock the measurement basis, a phase reference is generally required. However, directly introducing such a reference is incompatible with low-noise quadrature measurement. In add-through configurations, the input--output boundary condition is given by \cite{2016_microcavity_quantum_couple_theo}
\begin{equation}
\hat a_{j,\mathrm{out}}(t)
=
\hat a_{j,\mathrm{in}}(t)
-
\sqrt{\kappa_{j,\mathrm{ex}}}\,\hat a_j(t),
\label{eq:io}
\end{equation}
where $\hat a_j(t)$ is the intracavity annihilation operator of mode $j$, $\kappa_{j,\mathrm{ex}}$ is the external coupling rate, and $\hat a_{j,\mathrm{in}}(t)$ and $\hat a_{j,\mathrm{out}}(t)$ denote the input and output fields at the coupling port in the pump polarization. This relation shows that any classical field component (or technical noise) injected through $\hat a_{j,\mathrm{in}}$ at the quantum sideband frequencies is directly transferred to the detected output, thereby elevating the measured noise floor and masking two-mode squeezed vacuum states.

\begin{figure*}[t]
    \centering
    \includegraphics[width=0.9\textwidth]{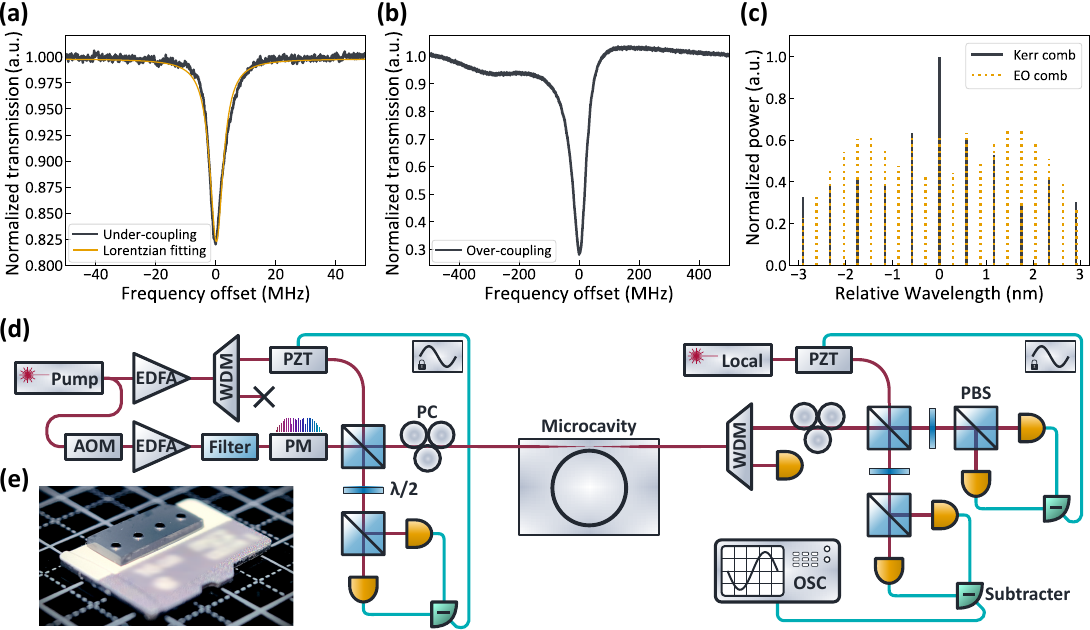}
    \caption{Device characterization and experimental implementation. (a) Transmission spectrum of the selected pump mode at under-coupling conditions. (b) Transmission spectrum under overcoupling, exhibiting a pronounced non-Lorentzian (Fano-like) lineshape. (c) Comparison between chaotic microcombs generated above threshold and electro-optic (EO) combs used as the coherent reference. (d) Experimental setup implementing the pump-referenced phase-transfer scheme, including the EO-comb generation, polarization multiplexing, phase-locking, and balanced-homodyne detection. AOM, acousto-optic modulator; EDFA, erbium-doped fiber amplifier; PM, phase modulator; WDM, wavelength-division multiplexer; PZT, fiber piezoelectric stretcher; PBS, polarizing beam splitter; PC, polarization controller; OSC, oscilloscope. (e) Photograph of the microresonators shown next to a microSD card for scale.}
    \label{fig:setup}
\end{figure*}

In our experiment, we develop the CCC method as illustrated in Fig.~\ref{fig:concept}, where a weak electro-optic (EO) comb $\mathrm{R}_i$, serving as the reference field, is co-injected with the pump in an orthogonal polarization and frequency-shifted by an offset $\omega_{\mathrm{b}}$. The microcavity and its coupling are designed to avoid resonance with the reference field. As a result, the EO comb and the pump co-propagate along the same optical path, while the reference field remains separable from the quantum fields at the output and can therefore be detected independently to lock the measurement basis. Owing to the orthogonal polarization, frequency offset, and mode-selective design, the reference field is strongly rejected from the analyzed quantum channel. At the operating reference power, no resolvable reference-induced noise-floor elevation is observed in the balanced-homodyne readout. Thus, the CCC method  establishes a stable quadrature measurement basis that enables long-duration readout of audio-frequency signals. Even when quantum modes at different frequencies are distributed across separate locations, it provides a non-invasive phase reference for each mode, thereby supporting the practical deployment of on-chip CV quantum information processing.

\begin{figure*}[t]
    \centering
    \includegraphics[width=0.9\textwidth]{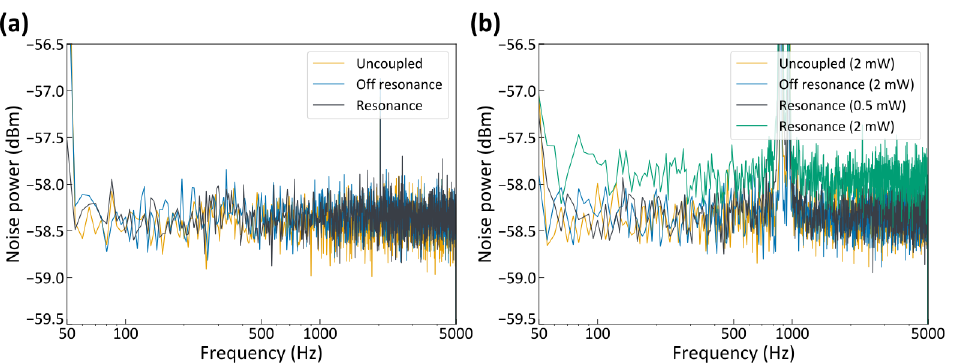}
    \caption{Reference-induced measurement-noise characterization. (a) Noise spectra measured without the coherent reference for three conditions (uncoupled, coupled off resonance, and coupled on resonance), showing that no noise-floor elevation is resolved within the measurement setup. (b) Noise spectra measured with the coherent reference under different coupling/resonance conditions: uncoupled ($P_{\mathrm{ref}}=2~\mathrm{mW}$, before the microcavity), coupled off resonance ($P_{\mathrm{ref}}=2~\mathrm{mW}$), and coupled on resonance at $P_{\mathrm{ref}}=0.5~\mathrm{mW}$ and $2~\mathrm{mW}$.  The narrow features around 800--1000~Hz originate from oscilloscope oversampling of the control signal and do not reflect the optical noise floor.}
    \label{fig:result1}
\end{figure*}

\section{The experiment}\label{sec:exp}

In the experiment, we use a chip-based silica microtoroid resonator fabricated based on the previous work \cite{2003_silica_reflow_vahala,2017_silica_reflow_jiang}. The diameter of the fabricated microresonator is 732 $\upmu\mathrm{m}$ with a free spectral range (FSR) of approximately 72.8 GHz.  The intrinsic quality factor is measured to be approximately $4\times 10^7$ at under-coupling conditions, as shown in Fig.~\ref{fig:setup}a. A tapered fiber is attached to the microcavity to achieve overcoupling, which suppresses the effect of optical radiation pressure on the cavity's dynamic processes and leads to an escape efficiency of 64.3\%. Under overcoupling, the transmission spectrum exhibits a pronounced non-Lorentzian (Fano-like) lineshape, even after thermal effects are taken into account, as shown in Fig.~\ref{fig:setup}b. This behavior is consistent with interference between the cavity resonance and an additional background or parasitic channel in the loaded optical response. For the cavity-FSR calibration, we pump the cavity mode above threshold to generate chaotic microcombs. The $\pm1$-FSR lines of the chaotic microcomb are then aligned with the $ \pm 2$-order sidebands of an EO comb generated by deep modulation of the pump field without the frequency shift introduced by the acousto-optic modulator (AOM). By tuning the EO modulation frequency, this alignment gives $\omega_{\mathrm{cavity}}=2\omega_{\mathrm{EO}}$, with $\omega_{\mathrm{EO}}/2\uppi=36.4~\mathrm{GHz}$ and a cavity FSR of approximately $72.8~\mathrm{GHz}$, as shown in Fig.~\ref{fig:setup}c.

The experimental setup is provided in Fig.~\ref{fig:setup}d, where the pump field is split into two paths as described above. One part is shifted by 181.4~MHz using the AOM, amplified through an erbium-doped fiber amplifier (EDFA), and deeply modulated by a phase modulator at 36.4~GHz to generate the reference EO combs. The other part is amplified through an EDFA to serve as the pump field for generating the two-mode squeezed state. We employ wavelength-division multiplexers (WDMs) and a home-made traveling-wave filter cavity to suppress the amplified-spontaneous-emission fluorescence and technical noise introduced by the EDFAs; the filter positions and main parameters are summarized in Supplementary materials S3 online. The two light fields are then combined with orthogonal polarizations and injected into the microcavity through a tapered fiber with 94\% transmittance. During the squeezing measurements, the cavity is pumped below threshold at a pump wavelength of $1549.768~\mathrm{nm}$. The pump power before entering the microcavity is set to $P_{\mathrm{pump}}=54.2~\mathrm{mW}$, and the coherent reference power is set to $P_{\mathrm{ref}}=0.5~\mathrm{mW}$. A WDM is used to filter out the pump light after it exits the cavity. The bichromatic local light field is generated by two identical lasers at $1549.184~\mathrm{nm}$ and $1550.352~\mathrm{nm}$, respectively, matching the $\pm1$-FSR cavity modes relative to the pump. Thus, the balanced homodyne detector measures the two-mode squeezed quantum sideband pair $\mathrm{Q}_1$ and $\mathrm{Q}_2$ in Fig.~\ref{fig:concept}. The optical power of each local-light frequency component is set to $3.5~\mathrm{mW}$. We separate the polarization-orthogonal squeezed light and reference beams by a polarizing beam splitter (PBS) after the microcavity. The squeezed light is coupled with local light field by a  combination of half-wave plate and PBS, and detected by a balanced homodyne detector. The transmission efficiency of the detection chain (from the tapered fiber output to the input of the homodyne detector) is 66.0\%. The homodyne detection efficiency is $97.4\%$ (see Supplementary materials S3 online).

All quadrature measurements are performed with the homodyne phase stabilized by the pump-referenced phase lock, enabling stable access to a calibrated quadrature basis (see details in Supplementary materials S1 online). To normalize the quadrature values, the SNL is measured under the same local light powers and electronic settings as the squeezing measurements while blocking the microcavity output and keeping the dual-local-oscillator balanced-homodyne chain active. For the time-domain examples, we record representative quadrature-combination traces for the squeezed, anti-squeezed, and SNL settings without applying any signal filtering. We then estimate the corresponding spectra from the time-domain samples using Welch's method with a rectangular (boxcar) window, without applying any filtering.

To quantify entanglement, the two-mode covariance matrix is reconstructed from a set of directly measured quadrature combinations, rather than from a single simultaneous measurement of $(q_1,q_2,p_1,p_2)$. By switching the locking angle and the local-oscillator configuration (single-color and two-color settings), we acquire the photocurrents of the inter-mode combinations $q_1-q_2$, $p_1-p_2$, $q_1-p_2$, and $p_1-q_2$, together with the single-mode quadratures $q_1,p_1,q_2,p_2$. The photocurrents are filtered to suppress low-frequency technical noise below $60$~Hz and narrow spectral lines associated with the oversampling of the acquisition electronics, and the covariance elements and their statistical uncertainties are obtained from these records using standard variance relations and a moving-block bootstrap \cite{2016_bootstrap_symplectic+eigenvalue,2024_NTT_tomography_bootstrap}. The reconstructed covariance matrix is then used to evaluate entanglement via the positive partial transpose (PPT) criterion  \cite{2000_PPT_Simon_theo,2000_PPT_hbar2_theo}. Details of the reconstruction procedure and PPT criterion are provided in the Supplementary materials S2 online.

\section{The results}\label{sec:results}

\begin{figure*}[t]
    \centering
    \includegraphics[width=0.9\textwidth]{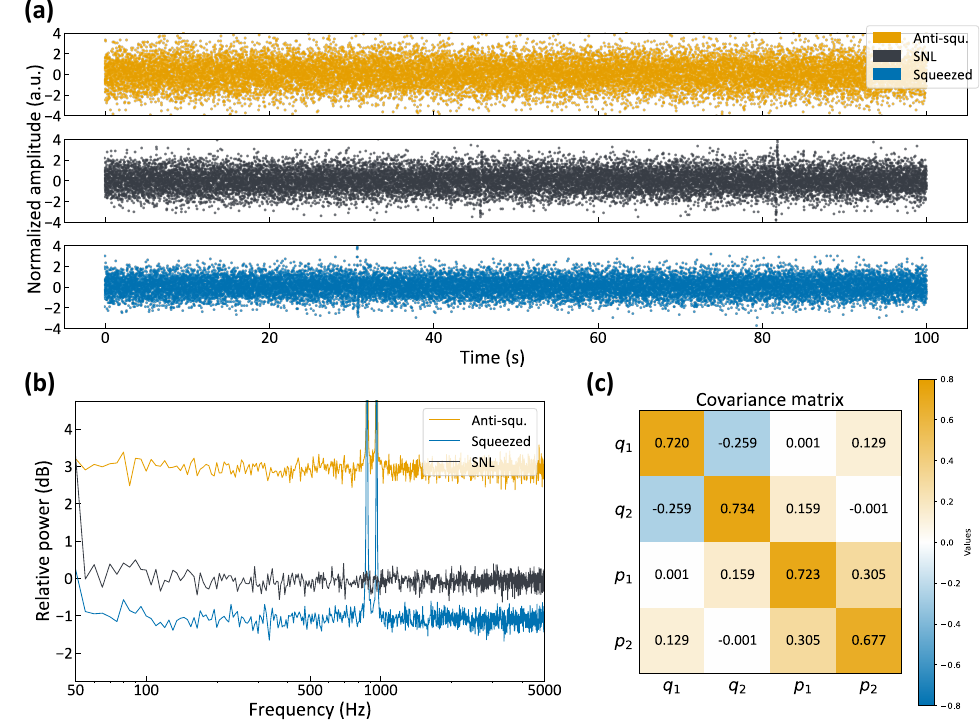}
    \caption{Phase-locked quadrature measurement, audio-band squeezing, and entanglement verification of the phase-stabilized two-mode squeezed state. (a) Representative time-domain quadrature fluctuations (anti-squeezed, SNL, and squeezed) recorded over a $100\,\mathrm{s}$ acquisition window with 10 kSa/s sampling rate after phase locking, showing stable access to different quadratures and stable locked operation during the measurement interval. (b) Frequency-domain squeezing and anti-squeezing spectra demonstrating observable squeezing down to the audio band. (c) Reconstructed covariance matrix. SNL, shot-noise level.}
    \label{fig:result2}
\end{figure*}

To evaluate the influence of the coherent reference light on the experimental system, we measure the output noise spectra with different reference powers. Specifically, we compare noise spectra recorded without the reference light and with reference light power of $P_{\mathrm{ref}}=0.5~\mathrm{mW}$ and $P_{\mathrm{ref}}=2~\mathrm{mW}$, as shown in Fig.~\ref{fig:result1}. Without the reference light, no observable noise-floor elevation is found for the uncoupled, off-resonance, and on-resonance conditions, which means that the noise of pump light does not contribute to the measured noise spectrum, as shown in Fig.~\ref{fig:result1}a.

With the reference light co-propagating with the pump light, the spectrum remains unchanged for the uncoupled and off-resonance cases at $P_{\mathrm{ref}}=2~\mathrm{mW}$, whereas a clear elevation is resolved only when the system is tuned on resonance. Importantly, this on-resonance elevation is reference-power dependent and as $P_{\mathrm{ref}}$ is reduced to $\sim0.5~\mathrm{mW}$, it eventually becomes indistinguishable from the off-resonance case, as shown in Fig.~\ref{fig:result1}b. Thus, we operate the system at $P_{\mathrm{ref}}=0.5~\mathrm{mW}$, which ensures the experimental system is not affected by the coherent reference light.

In the measurement, we phase-lock the homodyne detection to the pump-referenced phase reference with the CCC method, enabling stable switching between the squeezed and anti-squeezed combinations of quadratures. We show the measured squeezed, anti-squeezed, and SNL noise in the time domain with $\sim 10^6$ samples in $100~\mathrm{s}$, indicating stable locked operation over the acquisition window, as shown in Fig.~\ref{fig:result2}a. Using the Welch's method, we obtain the corresponding noise spectra in Fig.~\ref{fig:result2}b. Although the spectra in Fig.~\ref{fig:result2}b are displayed from $50~\mathrm{Hz}$, the calibrated SNL begins to rise appreciably below approximately $60~\mathrm{Hz}$. This behavior indicates that low-frequency classical and technical noise becomes non-negligible and that the readout is no longer demonstrably SNL in this region. Therefore, the apparent separation between the squeezed spectrum and the elevated SNL below $60~\mathrm{Hz}$ is not used as quantitative evidence of quantum-noise suppression. We conservatively use only analysis frequencies at or above $60~\mathrm{Hz}$ to support the reported squeezing, and frequency components below $60~\mathrm{Hz}$ are excluded from the covariance-matrix reconstruction during preprocessing. The $50~\mathrm{Hz}$ to $60~\mathrm{Hz}$ interval is retained in the figure only to display the behavior near the low-frequency boundary. The upper frequency of $5~\mathrm{kHz}$ is set by the $10~\mathrm{kSa/s}$ sampling rate. Within this quantitative analysis band, the spectra show $\sim$1.03~dB (inferred on-chip $1.73~\mathrm{dB}$) of two-mode squeezing below the SNL from $60~\mathrm{Hz}$ to $5~\mathrm{kHz}$. This result confirms the existence of quantum correlation between two modes, but it is not sufficient to characterize the existence of entanglement.

In order to verify the entanglement, we reconstruct the covariance matrix by switching the locking angle and local-oscillator configuration to measure a set of single- and inter-mode quadratures. The measured covariance matrix is shown in Fig.~\ref{fig:result2}c, which confirms entanglement via the PPT criterion with a minimum symplectic eigenvalue $\tilde{\nu}_{-}=0.395$ ($<0.5$). To quantify the statistical uncertainty of the minimum symplectic eigenvalue $\tilde{\nu}_{-}$ inferred from a finite-length record, we apply a moving-block bootstrap to the $100~\mathrm{s}$ time trace ($\sim10^6$ samples) to preserve temporal correlations in the homodyne readout. We generate 400 bootstrap replicates, for each bootstrap replicate we reconstruct $\boldsymbol{V}$ and recompute $\tilde{\nu}_{-}$, yielding $\tilde{\nu}_{-}=0.395\pm0.001$ (one standard deviation), with percentile confidence intervals $\tilde{\nu}_{-}\in[0.394,\,0.396]$ (68\%) and $\tilde{\nu}_{-}\in[0.393,\,0.397]$ (95\%). All intervals remain well below the PPT threshold, indicating that the entanglement certification is robust against finite-sample statistical fluctuations.

\section{Conclusion and discussion}\label{conclusion}

In summary, we demonstrate phase-referenced, on-chip two-mode squeezing in the audio-frequency band using the CCC method. By routing the sensing and actuation for phase stabilization through a weak, co-propagating EO reference comb that is orthogonally polarized and detected separately from the quantum modes, we stabilize the homodyne measurement basis without directly perturbing the squeezed modes, enabling deterministic quadrature access and covariance-matrix reconstruction. We observe a measured two-mode squeezing of $-1.03~\mathrm{dB}$ down to $60~\mathrm{Hz}$, corresponding to an inferred on-chip generated squeezing of $-1.73~\mathrm{dB}$ after correcting for the transmission and detection efficiencies. From the reconstructed covariance matrix, we verify entanglement via the PPT criterion, with a minimum symplectic eigenvalue $\tilde{\nu}_{-}=0.395\pm0.001$.

Although the observed squeezing level is only $\sim1.03$~dB in the audio band, our work presents the feasibility of generating on-chip audio-band squeezing. The squeezing level can be improved by increasing the escape efficiency and by reducing parasitic loss, mode mismatch, and unwanted background-channel coupling. Extending the observed squeezing below $60$~Hz will require further suppression of residual low-frequency technical noise on the channel-isolation, phase-control, and detection sides. On the channel-isolation side, leakage of reference and environmental noise into the quantum state can be further reduced by strengthening polarization isolation, lowering the technical noise of the EO comb, enhancing the microcavity's modal selectivity, and suppressing parasitic background channels that increase on-resonance coupling of auxiliary-field noise. On the phase-control side, operating-point drifts and parasitic interference can be suppressed by optimizing both the optical routing and the servo electronics. On the detection side, lower electronic noise floors can be achieved through improved electromagnetic shielding, enhanced thermal stability, and stronger acoustic isolation. With these improvements, the CCC method can support higher levels of audio-band squeezing below $60$~Hz. For a discussion of noise suppression, see details in Supplementary materials S3 online.

Our results pave the way toward practical deployment of audio-band, on-chip CV quantum technologies by addressing key challenges associated with low-frequency technical noise and phase-stable readout. The phase-stable, non-invasive control architecture is compatible with foundry-scale fabrication and packaging workflows and enables many-node, arrayed, distributed quantum sensing for field-relevant low-frequency measurements.

\section*{Conflict of interest}\vspace{-4pt}
The authors declare that they have no conflict of interest.

\section*{Acknowledgments}\vspace{-4pt}
This work was supported by the Innovation Program for Quantum Science and Technology (2024ZD0302403), the National Natural Science Foundation of China (U25A20525 and 12434015), and the Fund for Shanxi \textquotedblleft 1331 Project\textquotedblright\ Key Subjects Construction.

\section*{Author contributions}\vspace{-4pt}
Xiaolong Su conceived the original idea; Xuezhi Zhu and Meihong Wang performed theoretical analysis and calculations; Yaya He, Fan Zhang, and Xiaoshun Jiang fabricated the devices; Xuezhi Zhu, Yunyun Cao, and Rui Liu performed the experiment; Yaqing Zhang and Shiwei Du participated in part of the experiment; Xuezhi Zhu and Xiaolong Su analyzed the experimental data; and Xuezhi Zhu, Meihong Wang, and Xiaolong Su wrote the paper.

\section*{Appendix A. Supplementary material}\vspace{-4pt}
Supplementary materials to this article can be found online at link.


\bibliography{refs}

\end{document}